# Playing Fast Not Loose:
# Evaluating team-level pace of play in ice hockey using spatio-temporal possession data

David Yu*, Christopher Boucher, Luke Bornn, Mehrsan Javan

SPORTLOGiQ, Montreal, Quebec, Canada

*Email: david.yu@sportlogiq.com

## 1. Introduction

Pace of play is an important characteristic in ice hockey as well as other team-invasion sports. While in basketball pace has traditionally been defined as the number of possessions per 48 minutes, here we focus on pace and movement within a possession, leveraging the tremendous advancements in the capture of spatio-temporal data in team sports in recent years [1]. While much attention has been focused on speed and distance covered at the player level, spatio-temporal datasets also allow for more granular definitions of team-level pace of play such as measures of the speed between successive events or the speed of a possession as a whole.

While ice hockey has always been one of the fastest-moving sports, rule and tactical changes in the past 15 years, such as the removal of the rule limiting 2-line passes and the stricter enforcement of obstruction/holding infractions, have placed further emphasis on pace. At the start of the 2016-17 NHL season, Paul Maurice, head coach of the Winnipeg Jets said:

> "This game is just so fast now... I've seen fast players and I've seen fast teams, it's the first time I thought we had a fast league. The speed, to me, is the one thing that's changed more than anything. Our team, and the league as well, is as fast as I've ever seen it." [2]

Given the emphasis on pace in hockey in recent years, it is surprising that a recent study found a slight negative correlation between various metrics of forward attacking pace and offensive output such as shots and goals [3]. In this paper we not only explain this counterintuitive result, but also provide the first comprehensive study of pace within the sport of hockey, focusing on how teams and players impact pace in different regions of the ice, and the resultant effect on other aspects of the game. Our objectives are threefold:

1. Examine how pace of play varies across the surface of the rink, between different periods, in different manpower situations, between different professional leagues and rink surfaces, and through time between different seasons.
2. Determine how pace preceding various key events (such as shots, zone entries and passes) impacts their outcomes.

3. Quantify variations in pace of play at the team and player level and provide metrics to assess how well teams and players attack/defend pace.

Our results show that pace varies considerably, in both expected and unexpected ways, across all of the dimensions we examined and that pace is not strictly good or bad, but rather a delicate risk-reward balance.

## 2. Methods

### 2.1 - Dataset

We make use of SPORTLOGiQ's spatio-temporal dataset which has been used in a number of recent studies in ice hockey [3,4,5,6]. The dataset contains an average of ~3650 events per game with 21 primary event types and 89 distinct subtypes. Each event contains precise X,Y rink coordinates and timestamps. Furthermore, each event is labeled with the possession state making it easy to determine which team was in possession at the time of the event. Analyses were performed on all regular season National Hockey League (NHL), American Hockey League (AHL) and Swedish Hockey League (SHL) games in the 2016-17 and 2017-18 seasons. For analyses on the 2018-19 season, all regular season games up to and including November 24, 2018 have been included.

### 2.2 - Metrics of Pace

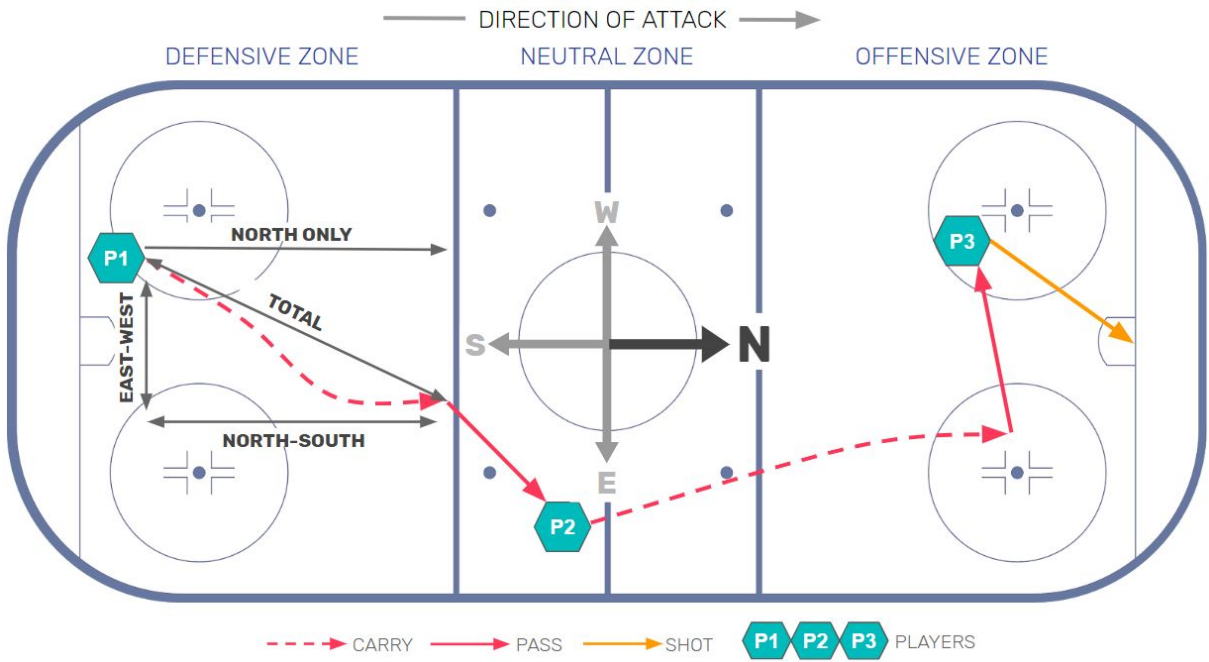

Figure 1 - Example possession sequence illustrating how pace is calculated in this study. P1 recovers puck in the DZ, carries to their DZ blueline and makes an outlet pass to P2 who carries into the OZ before passing to P3 for a one-timer shot. The shot is saved and held for a face-off thereby terminating the possession sequence. All events in this sequence have positive $\phi_N$ except for the pass from P2 to P3 which has a $\phi_N$ of zero.

We have used the distance travelled and time elapsed between successive possession events by the same team (i.e. passes, receptions, puck recoveries) to calculate various definitions of team-level pace. This includes total speed ($\phi_T$), as well as the east-west ($\phi_{EW}$), north-south ($\phi_{NS}$), and north-only ($\phi_N$) components of speed. We use conventional hockey terminology in defining directions where north is the direction of attack, and east-west represents play across the width of the rink (Figure 1). $\phi_N$ differs from $\phi_{NS}$ in that only forward progress is measured and any backward progress is assigned a $\phi_N$ of zero.

Our standard termination criteria was to end possession sequences when either the team in possession changed, the manpower situation changed, or a stoppage in play occurred. In these cases, the last event in the sequence was not included in the calculation of pace. This definition of pace means that possession sequences are allowed to continue even if the defending team makes a successful defensive play (e.g. blocked shot or save) so long as the play is not stopped and the attacking team regains possession of the subsequent loose puck.

### 2.3 - Zonal Analysis

Possessions were broken into sequences that occurred in each of the three zones (offensive, neutral, defensive). In addition to our standard termination criteria, possessions were also terminated when play transitioned between zones while the same team maintained possession. When this occurred, pace from the final event in the preceding sequence was assigned to the next sequence and the final event was then set as the first event of the subsequent possession sequence.

### 2.4 - Spatial Polygrid Analysis

We divided the rink into 668 equal sections measuring 5 ft. x 5 ft., which we term a polygrid (portmanteau of polygon grid). We then assigned the distance travelled and time elapsed between successive possession events equally to all cells that intersect the path between successive possession events. Only the standard termination conditions were used in this analysis.

Differential polygrids were made by aligning and subtracting speed values between two polygrids. In cases where limited sample size produced higher levels of noise in the differential polygrid, a 2D Gaussian kernel ($\sigma$ = 0.5) was applied to smooth the data prior to calculating the difference. This technique replaces each cell in the polygrid with a weighted average of itself and its neighbors. This was done for team-level polygrids when comparing team performance relative to league average.

### 2.5 - Team Level Analyses

Team-level analyses of attacking and defending pace were done using both the zonal and polygrid approaches. Metrics were only calculated at even-strength 5v5 for the NHL. Team attacking metrics are calculated for when a team is in possession, while team defending metrics measure the pace of the opposing team while a given team is defending.

## 2.6 - Player Level Analyses

Player-level analyses were calculated at even-strength 5v5 for the NHL using only the zonal approach due to smaller sample sizes. Players had to have played a minimum of 200 minutes at even-strength 5v5 to be included. Two different metrics were calculated at the player level:

### 2.6.1 - Individual Player Pace

Individual player pace was calculated by looking at only possession events a player directly participated in. For successive possession events, the distance and time components are assigned equally between the players associated with the two events. For example, a pass-reception sequence would be divided equally between the passer and receiver while a reception-shot by the same player would have the distance and time of the intervening carry assigned entirely to that player.

### 2.6.2 - With or Without You (WOWY) Plus Minus

The "with player" metric was calculated by averaging the team's attacking pace of all possession sequences while that player was on the ice. The "without player" metric was calculated by averaging the team's attacking pace while the player was not on the ice. The latter was done only for games where a player was in the lineup to better account for players that did not play all games with a given team in a season.

## 3. Exploring Pace

### 3.1 - Pace of Play by Zone

We first determined how pace varies between the offensive (OZ), neutral (NZ) and defensive (DZ) zones in the NHL for the 2017-18 regular season.

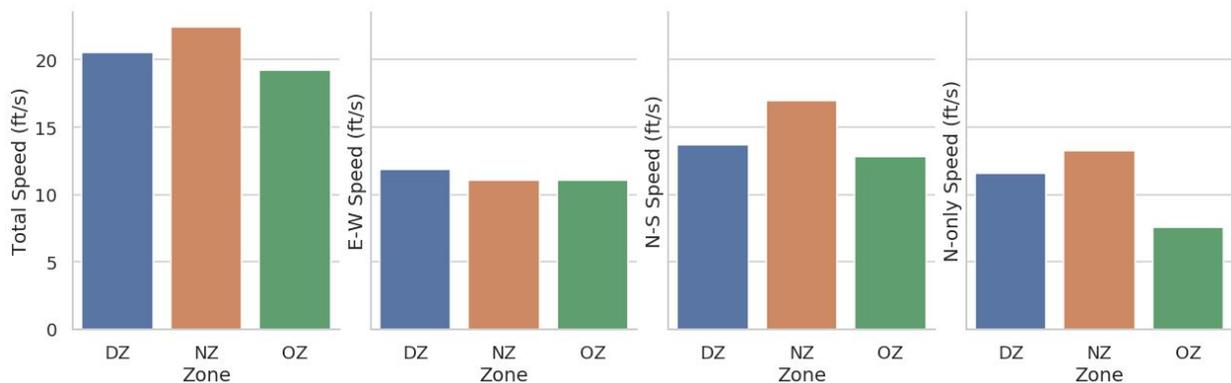

Figure 2 - Speed by Zone in the NHL for the 2017-18 regular season at even-strength (5v5)

We found that $\phi_T$ is highest in the NZ and generally ~10-13% slower in the OZ and DZ. The higher total speed in the NZ is driven largely by differences in $\phi_{NS}$ as $\phi_{EW}$ speed is roughly equal in all zones. Pace of play is relatively similar between the OZ and DZ across all directions with the exception of $\phi_N$. We find that $\phi_N$ in the OZ is 35% slower than DZ $\phi_N$ and 43% slower than NZ $\phi_N$.

Our analysis helps to explain some of the counterintuitive results obtained in prior studies. These studies have found that $\phi_N$ (also referred to as forward attacking or direct pace) displays a weak negative correlation with both offensive outputs such as shots and goals in hockey [3] as well as with team quality in English Premier League soccer [7, 8]. We believe this negative correlation is due to the large decline in $\phi_N$ as play enters the offensive zone since that's where shots and goals are generated and is where good teams spend proportionately more of their time.

Furthermore, the decline in $\phi_N$ in the offensive zone should be expected since there are diminishing returns for advancing forward in both ice hockey and soccer. In both sports, advancing forward along the sides of the playing surface leads to progressively worse shooting angles on net. In addition, both sports further disincentivize teams from advancing the puck/ball beyond the goal line. In hockey, this results in the puck being located behind the net while in soccer, this results in a turnover for the team in possession.

### 3.2 - Pace of Play by League (5v5)

We next examined how pace varies between the NHL and two of the top professional leagues in the world, the AHL and SHL.

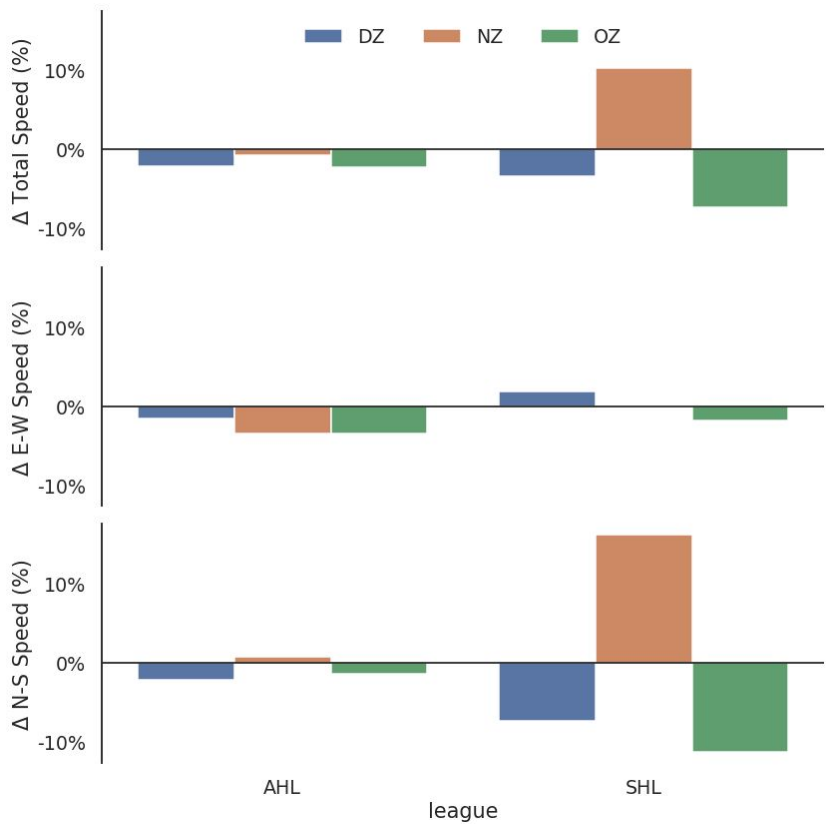

Figure 3 - League speed relative to the NHL for the 2017-18 regular season at even-strength (5v5)

The AHL is based in North America and serves as the primary developmental league for the NHL and is typically ranked as 4th or 5th best league in the world after the NHL [9]. AHL games are played on the same sized rink as the NHL (200 ft. x 85 ft.). Our analysis shows that $\phi_T$ is 1-2% slower compared to the NHL. The slowdown is more apparent in the OZ and DZ and is primarily driven by a decline in $\phi_{EW}$ across the three zones. This is likely due to the slightly lower talent levels in the AHL compared to the NHL. Since more talented players tend to play the puck more east-west and are able to play at higher speeds in the DZ and OZ where defensive pressure is generally higher.

The SHL is the highest division in Swedish ice hockey and is typically ranked as 3rd best league in the world after the NHL [9]. SHL games are played on international ice rinks that are considerably wider (~13.5 ft. or 16%), have longer neutral zones (~8 ft. or 16%) and a goal line much closer to the blueline (6 ft. or 9.4%) compared to North American rinks.

Pace in the SHL is considerably slower in both the DZ and OZ but is faster in the NZ. These changes are largely driven by changes in the $\phi_{NS}$ rather than $\phi_{EW}$ which remain relatively similar to the NHL. The higher NZ $\phi_{NS}$ in the SHL can likely be attributed to the much longer and wider NZ which limits the ability of defenders to apply NZ pressure and allows attackers to progress forwards relatively unchallenged. This is corroborated by the fact that the number of NZ passes per game in the SHL is 14% lower than the NHL and the lowest among the three professional leagues studied (Appendix Table 1).

In the DZ and OZ, comparably lower defensive pressure in the SHL caused by the extra width allows players to carry the puck more. This effectively lowers the $\phi_T$ since quick passes are not required for maintaining possession like they are in the NHL and AHL. This is supported by the fact that the SHL leads all leagues in OZ and DZ puck on stick possession time (Appendix Table 1).

Taken together the large differences in pace we see between NHL/AHL and SHL are likely due to the different rink dimensions favouring differing styles of play rather than differences in player ability. These large differences in pace between hockey played on International surfaces and North American surfaces likely contribute to the adjustment time needed for players transitioning between European and North American professional leagues.

### 3.3 - Pace of Play by Season (NHL 5v5)

Three years ago, the head coach of the Winnipeg Jets stated that the NHL game is as fast as he's ever seen it. It's worth examining how pace of play has changed since that time and whether pace of play has continued to increase.

Relative to the 2016-17 season, pace has increased slightly in the NHL over the last three years with largest increases occurring in the NZ (Figure 4). The increase in the NZ is driven more by an increase in $\phi_{EW}$ along with a more modest increase in $\phi_{NS}$. Higher $\phi_{EW}$ is likely driven in part by a 3-4% increase in east-west passes in the NZ over that same period (Appendix Table 2).

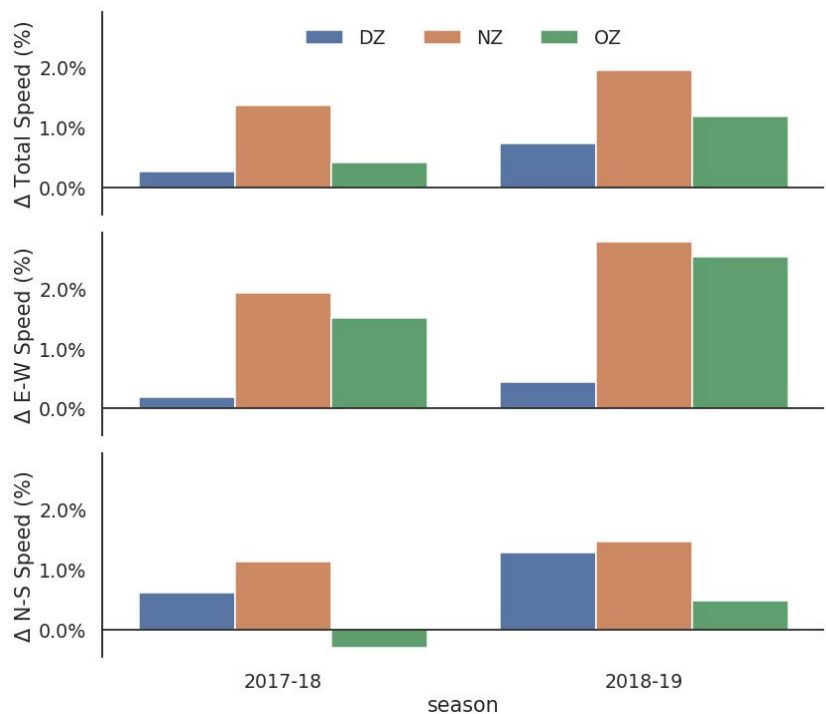

Figure 4 - Pace of play across seasons relative to the 2016-17 NHL regular season at even-strength (5v5)

### 3.4 - Pace of Play by Period

We also explored how pace varies between periods in regulation time. In ice hockey, the two team benches are located on either side of the red line and teams change ends after every period. This creates a situation where a team defends the goal closer to their bench in the 1st and 3rd periods and further from their bench in the 2nd period. The 'long change' in the 2nd period has been shown to increase goal scoring rates with the proposed rationale being that the long change makes it more difficult for tired defenders stuck in their own DZ to change [10, 11].

We don't see evidence for the 'tired defenders' hypothesis in our analysis of pace at 5v5 (Figure 5). Total speed in the OZ is actually lowest in the 2nd period, not what we'd expect if teams with possession in the OZ are taking advantage of tired defenders. Rather, we see an uptick in DZ pace in the 2nd period driven primarily by a ~7% increase in $\phi_N$. We believe this increase in DZ $\phi_N$ in the 2nd period is due to teams moving the puck forward quickly to either catch opposing teams off guard on bad changes or preventing them from changing all together. Indeed, we find the 2nd period has the lowest numbers of east-west DZ passes and controlled exits, and the highest number of forward stretch passes, which is likely responsible for the lower $\phi_{EW}$ and higher $\phi_N$ in the DZ (Appendix Table 3). Higher DZ and NZ $Φ_N$ in the 2nd period leads to a ~35% increase in odd man rushes (1-on-0, 2-on-1, 3-on-1, 3-on-2) which likely contributes to the increased scoring rates (Appendix Table 3).

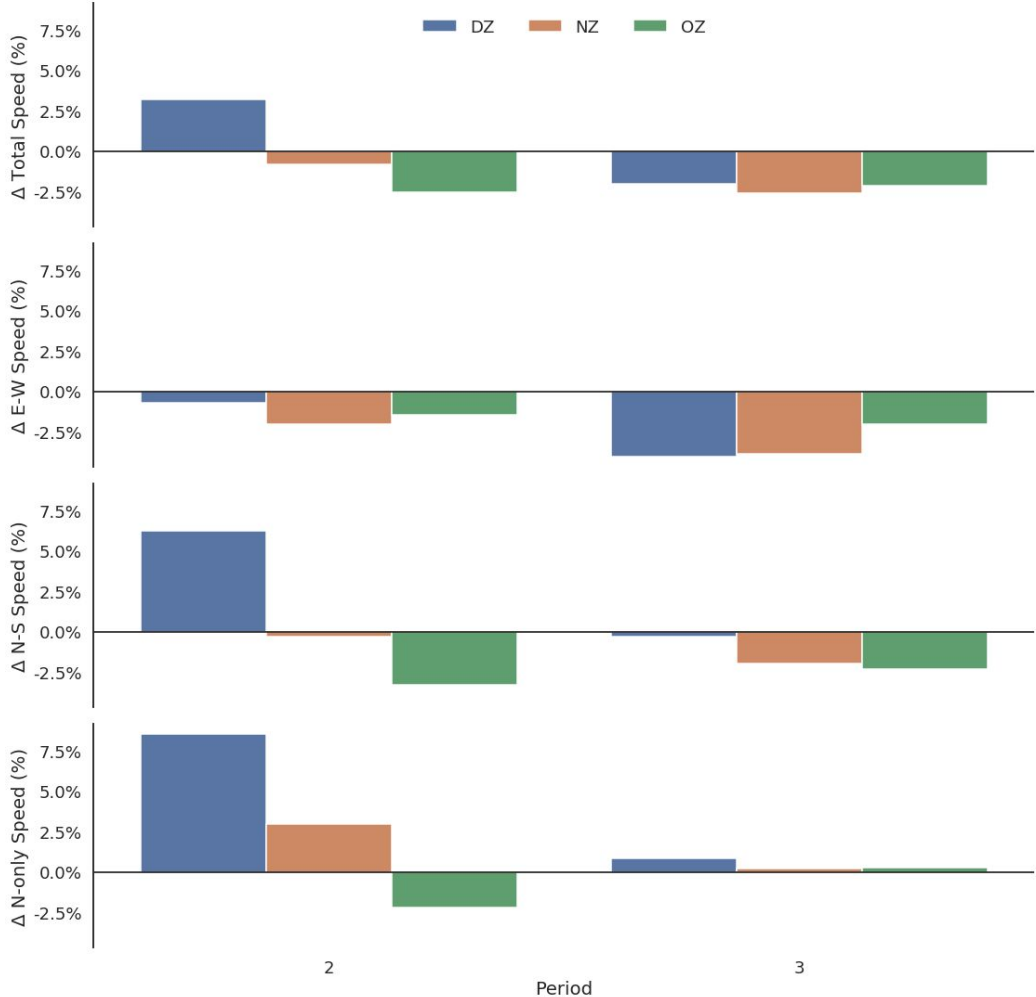

Figure 5 - Pace of play between periods in the 2017-18 NHL regular season at even-strength (5v5). Periods 2 and 3 are benchmarked relative to Period 1.

We also observe a progressive decline in $Φ_{EW}$ in all three zones from the 1st to 3rd periods. This effect is observed even after adjusting for close or tied score differentials (data not shown). We believe the slight decline in $Φ_{EW}$ may be attributed to more cautious, risk-averse play as the game progresses though this hypothesis bears further examination.

### 3.5 - Pace of Play by Manpower Situation

We next examined how pace varies across different manpower situations in the NHL (Figure 6). Total speed is lower in all zones at 4v4 though the increase in $Φ_{NS}$ in the NZ with corresponding decreases in the OZ and DZ is similar to the SHL and may be caused by a reduction in defensive pressure in all zones due to decreased player density.

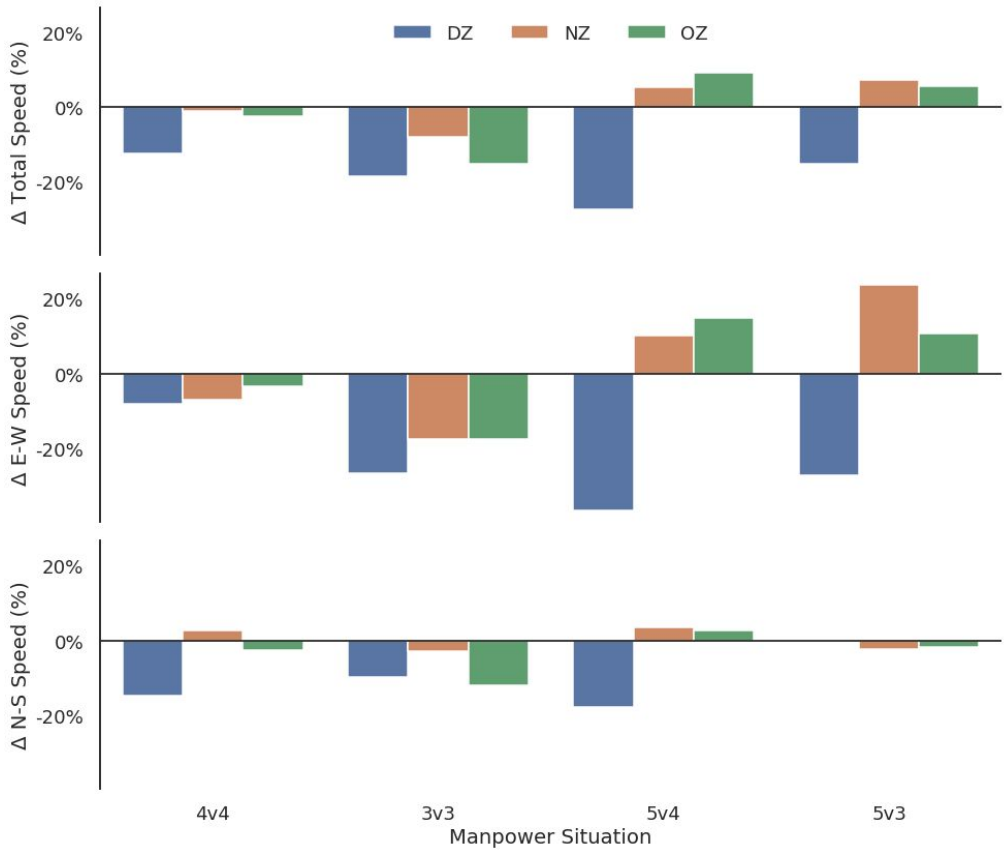

Figure 6 - Pace of play at different manpowers relative to even-strength
(5v5) for the 2017-18 NHL regular season

At 3v3, all forms of pace are slower across all zones. In the NHL, 3v3 is played only in sudden-death overtime and teams are typically deliberately slowing down and playing more cautiously since turnovers can often lead to a high danger counterattack for the opposing team.

On the powerplay at 5v4 and 5v3, we observe a large decline in DZ pace consistent with the slowing down of play as the team on the powerplay regroups after the opposing team clears their zone. In the NZ and OZ, pace is faster on the powerplay driven largely be an increase in $\phi_{EW}$ as teams try to break down defenses and draw the goalie out of position with cross-ice passes.

### 3.6 - Pace of Play Across the Rink (Polygrid)

To obtain a more granular view into how pace varies across the rink, we divided the rink into 668 equal sections measuring 5ft x 5ft. We then assigned the distance travelled and time elapsed between successive possession events equally to all grid sections on the path between these events. We used the polygrid approach to examine how various metrics of pace vary between 5v5 and 5v4 manpower situations (Figure 7).

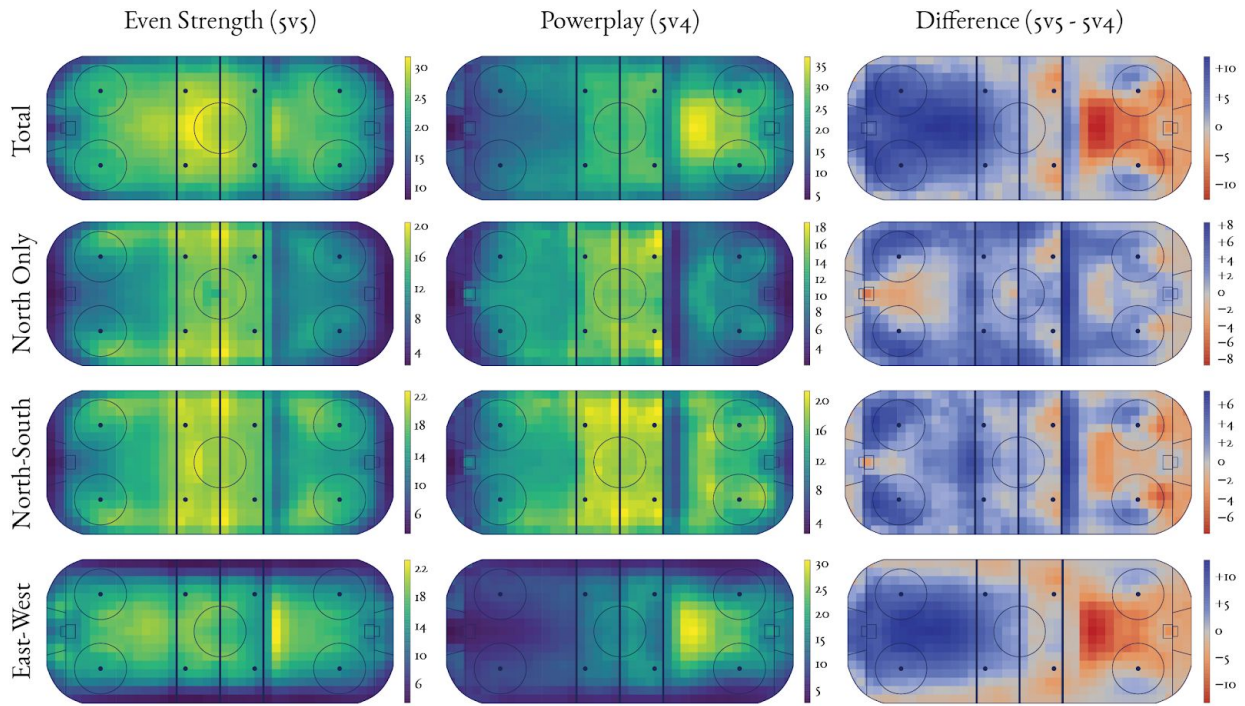

Figure 7 - Pace of play across the surface of the rink in the 2017-18 NHL regular season. Left: Even-strength (5v5) Center: Powerplay (5v4) Right: 5v5 - 5v4 difference (blue: faster at 5v5; red: faster at 5v4). All units are in ft/s.

The results show that pace is non-uniformly distributed across the length and width of the rink. For example, the effect of hockey's offside rule can be clearly seen at the offensive blue line with a marked decline in both $\phi_N$ and $\phi_{NS}$ and a peak in $\phi_{EW}$. We also measured differences in pace between even-strength (5v5) and power play (5v4) situations. While pace on the power play increases in large sections of the offensive zone (red), it declines by a similar magnitude (blue) in the defensive half. Results from the polygrid analysis are consistent with the differences between 5v5 and 5v4 speed shown in the zonal analysis but provide a much more granular view of how pace varies within each zone.

## 4. Impact of Pace

### 4.1 - Pace Preceding Zone Entries

Past research in ice hockey has shown that controlled entries into the OZ result in more favourable outcomes than dump-in entries [12]. However, even among controlled entries, not all are of equal value. SPORTLOGiQ event data tracks the skater differentials for every controlled entry. Here we used the percentage of entries with a shot on goal after as well as the shooting percentage of shots taken within 5 seconds following a zone entry to classify entries into high, mid, low, and very low danger (Table 1). Notice that not all odd man rushes (where attackers outnumber defenders) are high danger with 3-on-2 entries having a far lower chance of scoring than other types of odd man entries. We calculated $\phi_T$ of all events in the possession sequence preceding an entry and found that higher danger entries occur at a higher pace. $\phi_T$ preceding high danger entries is approximately 13% faster than that preceding dump-ins. The data presented is for the 2017-18 NHL regular season but the results are applicable to the AHL and SHL (Appendix Table 4).

| Entry Type | Shot after Entry % | Shooting % | Entry Class | $\phi_T$ (ft/s) |
|---|---|---|---|---|
| 1-on-0 | 66.6% | 26.2% | High Danger | 24.3 |
| 3-on-1 | 48.8% | 25.5% | | |
| 2-on-1 | 43.7% | 22.0% | | |
| 3-on-2 | 31.5% | 10.4% | Medium Danger | 23.4 |
| 1-on-1 | 29.9% | 8.8% | | |
| 2-on-2 | 26.2% | 7.1% | | |
| 3-on-3 | 20.8% | 5.2% | Low Danger | 22.5 |
| 1-on-2 | 21.3% | 4.8% | | |
| 2-on-3 | 21.6% | 4.6% | | |
| dump-in | 1.1% | 6.2% | Very Low Danger | 21.6 |

Table 1 - Pace of play preceding zone entries at even-strength (5v5) in the NHL for the 2017-18 regular season.

### 4.2 - Pace Preceding Shots

We measured $\phi_T$ in the 5 seconds. preceding non-deflected shot attempts at even-strength (5v5) in the 2017-18 NHL regular season (Figure 8). We excluded all deflected shots due to the inherent randomness of outcomes resulting from deflections. Shots were divided into quintiles by the pre-shot $\phi_T$. Average pre-shot speed varied from 10 ft/s to 42 ft/s between the lowest and highest $\phi_T$ quintiles. True shooting percentage, which is defined as the number of goals divided by the total shot attempts, increases from 2.9%-4.1%, an increase of 38% comparing the lowest to highest $\phi_T$ quintiles. Average shot distance remains fairly constant between $\phi_T$ quintiles (37-40 ft.) suggesting that the pace of pre-shot movement, and not shot distance, is the determining factor in increased shot quality.

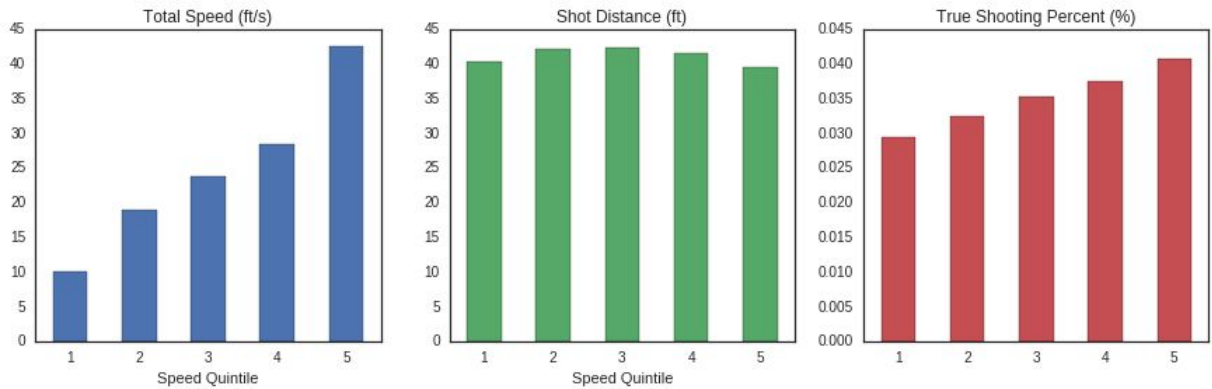

Figure 8 - Effect of total speed preceding a shot on shot distance as well as the shooting percentage in the 2017-18 NHL regular season (5v5). True shooting percentage is calculated as goals divided by total shot attempts.

These results are generally applicable in the AHL and SHL (Appendix Table 5). Since pace has been shown to improve to shot quality independently of shot location, we believe that expected goals models should incorporate pre-shot pace as a feature.

### 4.3 - Effect of Pass Speed on Reception Success

SPORTLOGiQ event data contains the coordinates and timestamps for all passes and receptions. Receptions can be classified as failed if the pass touches the receiver's blade but they fail to gain possession. We examined the effect of pass speed on reception outcome for all even-strength 5v5 passes in the NHL during the 2017-18 NHL season (0.50 million). We used SPORTLOGiQ pass types to account for variability in pass length, angle and difficulty. Aside from passes to the slot, failed receptions from all other pass types occur at significantly higher speeds than successful receptions (Figure 9).

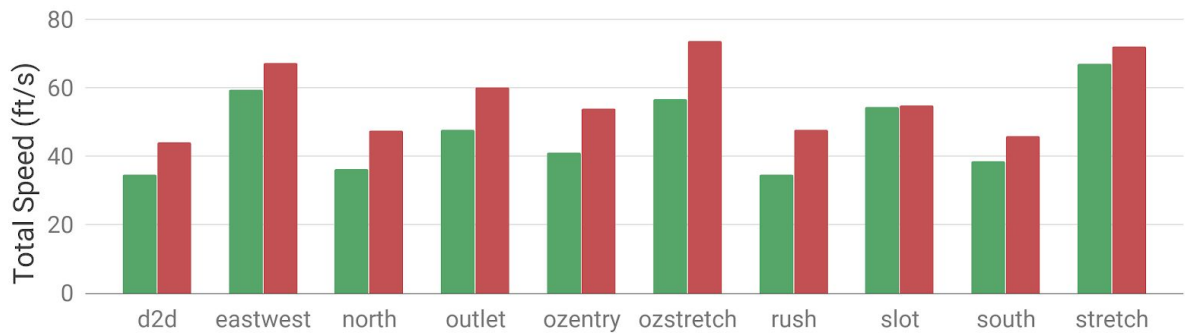

Figure 9 - Effect of pass speeds on reception outcomes for various pass types. Successful receptions are green while failed receptions are red.

## 5. Team and Player Effects

### 5.1 - Team-level Pace (Zonal)

We examined team-level pace using both the zonal and polygrid methods. We benchmarked a given team's attacking and defending pace relative to the league average. Analyses were performed for the 2016-17 and 2017-18 NHL regular season (Figure 10).

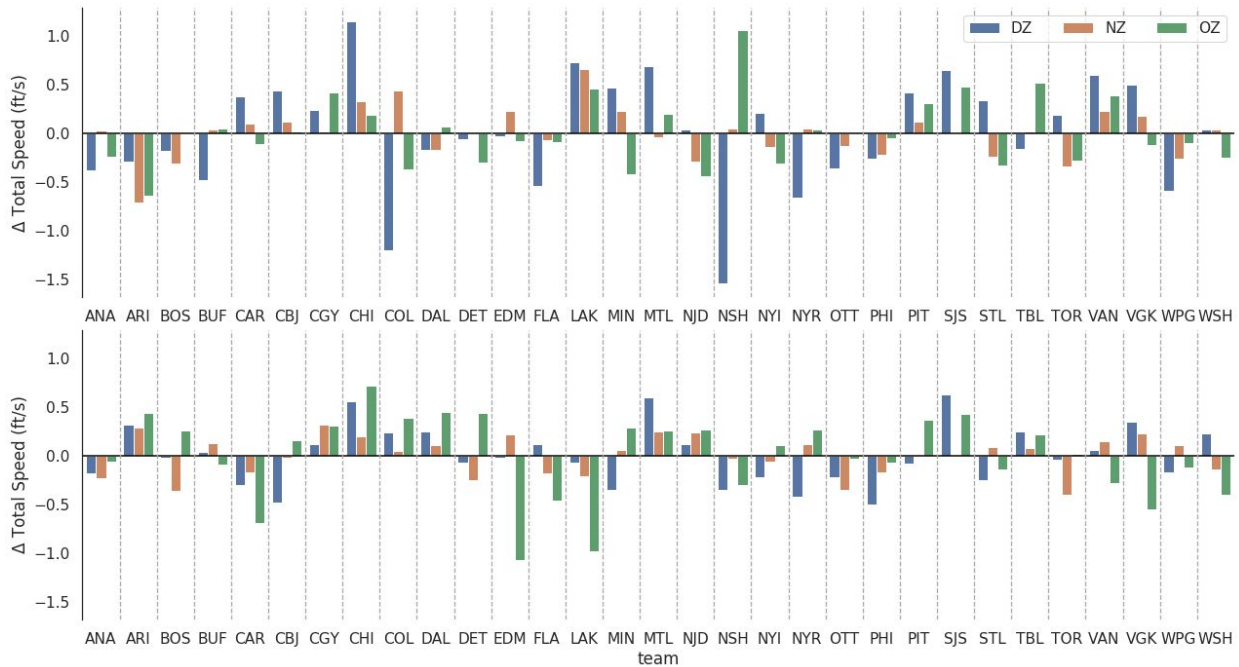

Figure 10 - Zonal analysis of total speed by team in the 2017-18 NHL regular season. Top: Team Attacking vs. NHL Average. Bottom: Team Defending vs. NHL Average

Zonal analysis shows that teams vary in their abilities to attack or defend pace in different zones. While attacking, teams like Chicago (CHI) and Los Angeles (LAK) were consistently faster than league average in all three zones while others like Arizona (ARI) and Winnipeg (WPG) were consistently slower. These teams are the exception since most teams were faster in some zones and slower in others. For example, Nashville (NSH) had the lowest DZ $\phi_T$ and the highest OZ $\phi_T$ of any team in the NHL.

While defending, some teams consistently gave up more (e.g. ARI, CHI, MTL) or less (CAR, LAK) pace through all zones though once again, most teams display variability between different zones. Differences in team attacking and defending speed relative to league average are somewhat repeatable across seasons in the NHL (Appendix Figures 1 and 2)

### 5.2 - Team-level Pace (Polygrid)

We also explored variation in pace at the team-level using the polygrid approach. This allowed us to obtain a more granular view of what areas of the ice a team is most or least effective at generating or preventing pace (Figure 11).

While the overall results of the polygrid analysis largely corroborates those found in the zonal analysis, we are able to discern patterns within zones that would otherwise be missed. For example, while attacking, some teams show asymmetry in DZ $\phi_T$ with faster pace on either the left (LAK, VAN) or right (CAR, CHI) side of the ice. Some teams are also much slower (NSH, ANA) or faster (CHI, VAN) around their own net in the DZ and this correlates well with the team's tendency to perform controlled breakouts (NSH - 1st; ANA - 5th; CHI - 28th; VAN - 30th) which slow down the pace of DZ exits.

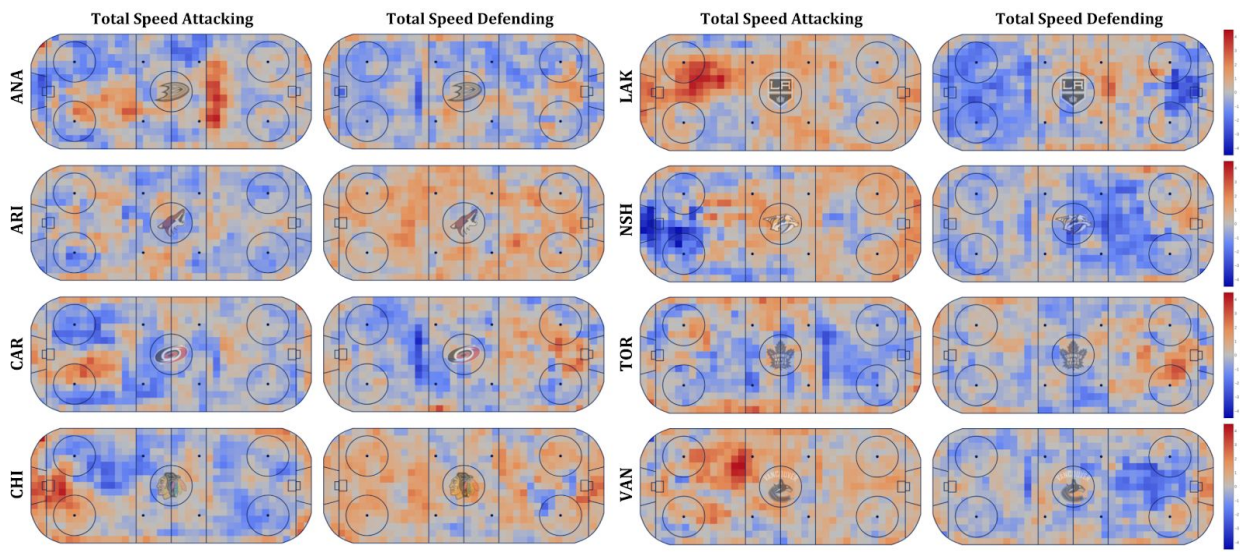

Figure 11 - Polygrid analysis of total speed by team in the 2017-18 NHL Season. Left hand side of the rink is the team's DZ and right hand side is the team's OZ for both attacking and defending polygrids. All units are in ft/s.

There is also considerable variation in attacking pace within the OZ where some teams have much higher (ANA) or lower (TOR) pace along the OZ blueline. Pace in this region of the ice is primarily due to possession maintaining EW passes between defenceman and likely does little to contribute to higher danger scoring chances. Some teams are also faster along the boards (CHI, TOR) while others appear to play with higher pace throughout the entire OZ (NSH, LAK).

On the defensive side, teams also exhibit differences in where they are most effective at slowing down their opponents. For example, LAK is very effective at slowing down teams around the opposing team's net, then give up pace through most of the NZ, but then effectively slow down pace across large swathes of their own DZ.

We believe that team-level differential polygrids are a useful way to visualize attacking and defending tendencies. Differential polygrids for all remaining NHL teams for the 2017-18 season can be found in Appendix Figure 3.

### 5.3 - Player-level Pace (Zonal)

At the player level we chose to examine both individual player pace (for possession events that player was directly involved in) as well as WOWY plus-minus (on/off splits of team-level attacking pace) using the zonal approach (Figure 12). For individual speed, we noticed that defenceman were much slower than forwards in the DZ while the opposite was true in the OZ (Appendix Figure 4). This discrepancy is likely due to the varying amounts of defensive pressure applied by the opposing team on forwards and defenceman in these two zones. As such, individual player pace metrics were adjusted for team, position, and zone for all analyses. Adjusting for team-differences was done to bring individual player pace analyses in line with WOWY (which by definition is team-adjusted) though team-adjusted metrics will penalize players playing on faster teams and vice versa.

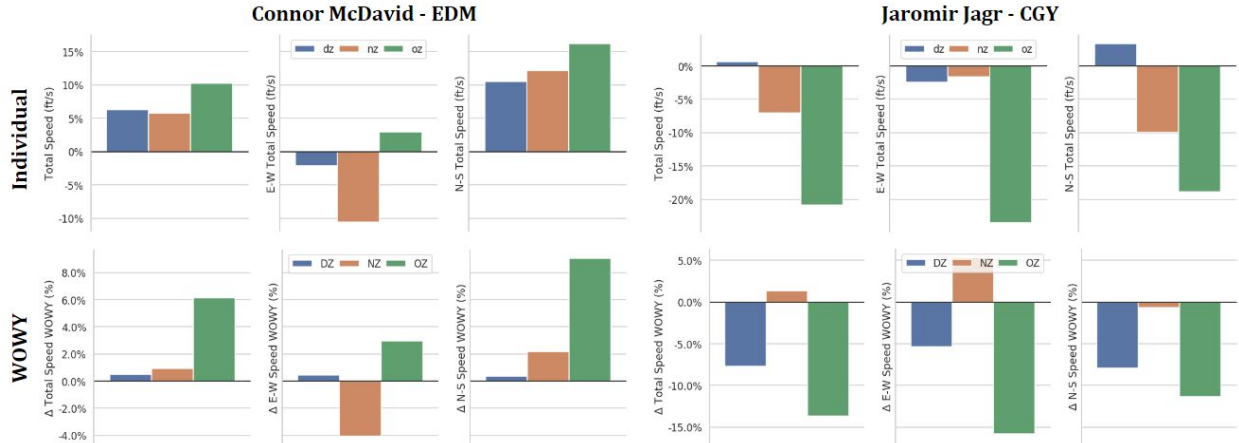

Figure 12 - Comparison of Individual and WOWY speed for Connor McDavid (EDM) and Jaromir Jagr (CGY) in the 2017-18 NHL regular season.

Connor McDavid is considered by many to be the fastest and most dynamic player in the NHL today [13]. Jaromir Jagr is one of the all-time greats, having accumulated the 2nd most career points in the NHL after Wayne Gretzky. However, Jagr was 45 years old in the 2017-18 NHL season making him by far the oldest player in a league getting younger and faster every year. Examining pace for these two forwards using both the Individual and WOWY methods shows McDavid to be one of the fastest and Jagr to be the slowest player in the NHL in the OZ. There is good correspondence between the Individual and WOWY metrics, though the differences are typically magnified for individual pace compared to the WOWY pace. This is to be expected since individual pace should be less affected by the quality of your line-mates compared to WOWY pace.

Comparisons of $\phi_T$ to the $\phi_{EW}$ and $\phi_{NS}$ components show that Connor McDavid drives pace primarily through higher speeds in the north-south direction which is consistent with the evaluation of most hockey experts [13]. The top and bottom 20 players for each zone for the 2017-18 NHL regular season, ranked by total speed WOWY difference, are given in the (Appendix Tables 6-11)

## 6. Discussion

We provide the first comprehensive review of team-level pace in ice hockey showing how pace varies in different areas of the ice, between leagues (and rink surfaces), across seasons, between periods, and when manpower situations change. Furthermore, we demonstrate how pace impacts the outcomes of key events. Our findings suggest that increased team-level pace is beneficial, but perhaps only up to a certain point. Higher pace can create breakdowns in defensive structure and lead to both higher danger zone entries and improved shot quality. On the other hand, our pass reception analysis shows that very high pass speeds can lead to more turnovers. Finally, we show that teams and players vary in their ability to attack and defend pace in different zones and areas of the ice surface. Future work at the player-level will use an adjusted plus-minus model to account for the effects of teammates and opposition on a player's performance.

Our analysis also suggests that forward attacking pace ($\phi_N$), which is currently the most widely used metric of team-level pace in both hockey and soccer [3, 7, 8], is not an ideal metric for measuring either offensive output or team quality. This is because $\phi_N$ declines by a large amount as play progresses closer to the opponent's goal. $\phi_N$ may serve as a useful tool to gauge team-level pace in the defensive and neutral zones but we believe that $\phi_T$ or perhaps $\phi_{EW}$ are better metrics once play has entered the offensive zone or offensive third.

Taken together, our results demonstrate that measures of team-level pace derived from spatio-temporal event data are informative metrics in ice hockey and may prove useful in other team-invasion sports. Defining the pace of play as the speed of on the puck actions rather than a set of player trajectories is a better estimate of team-level pace as it implicitly captures game context. Using this definition, our approach can be easily extended to other sports and leagues where no player tracking data is available. Future work will explore team-level pace in other sports like soccer, basketball, rugby or handball to see if similar patterns exist.

## Acknowledgements

The authors would like to thank Yi Zhou for help with graphic design and Evin Keane, Nick Czuzoj-Shulman, Matt Perri, and Sam Gregory for helpful discussions and comments.

# Appendix

| League | Season | DZ Passes | DZ Time (min.) | NZ Passes | NZ Time (min.) | OZ Passes | OZ Time (min.) |
|---|---|---|---|---|---|---|---|
| NHL | 2017-18 | 352.5 | 12.6 | 115.0 | 5.4 | 250.0 | 8.9 |
| AHL | 2017-18 | 323.0 | 11.5 | 103.1 | 4.8 | 223.2 | 8.0 |
| SHL | 2017-18 | 327.6 | 12.7 | 98.5 | 4.8 | 239.0 | 9.2 |

Appendix Table 1 - Comparison of passing and possession time by zone in the NHL/AHL/SHL (even-strength - 5v5). Pass metrics include both successful and failed attempts and are averaged per game. Zone possession time metrics are also averaged per game.

| League | Season | Manpower | EW >10ft. Passes | EW >15ft. Passes |
|---|---|---|---|---|
| NHL | 2016-17 | 5v5 | 47.9 | 39.8 |
| NHL | 2017-18 | 5v5 | 49.1 | 40.7 |
| NHL | 2018-19 | 5v5 | 49.7 | 41.3 |

Appendix Table 2 - Neutral zone east-west passing tendency by season in the NHL (even-strength - 5v5). Successful passes with greater than 10 or 15ft of EW distance were counted. Passes must have both originated in and been received in the NZ. Metrics are averaged per 60 minutes.

| League | Season | Period | DZ Controlled Exits | DZ D2D Passes | DZ Stretch Passes | Odd Man Rushes |
|---|---|---|---|---|---|---|
| NHL | 2017-18 | 1 | 21.6 | 44.1 | 11.8 | 2.65 |
| NHL | 2017-18 | 2 | 18.6 | 37.3 | 12.7 | 3.49 |
| NHL | 2017-18 | 3 | 21.2 | 39.8 | 11.5 | 2.44 |
| AHL | 2017-18 | 1 | 19.9 | 41.7 | 10.3 | 2.84 |
| AHL | 2017-18 | 2 | 17.1 | 35 | 10.5 | 3.55 |
| AHL | 2017-18 | 3 | 19.1 | 37.1 | 9.8 | 2.52 |
| SHL | 2017-18 | 1 | 19.6 | 46 | 10.9 | 2.04 |
| SHL | 2017-18 | 2 | 16.5 | 38.7 | 12.4 | 2.57 |
| SHL | 2017-18 | 3 | 19.2 | 42 | 11.2 | 1.42 |

Appendix Table 3 - Defensive Zone Tendencies and Odd Man Rushes by Period. All metrics are reported as per game averages.

| League | AHL | | | SHL | | |
|---|---|---|---|---|---|---|
| Entry Type | Shot after Entry % | Shooting % | Speed (ft/s) | Shot after Entry % | Shooting % | Speed (ft/s) |
| 1-on-0 | 69.1% | 24.3% | 23.8 | 69.6% | 24.1% | 23.9 |
| 3-on-1 | 47.5% | 25.1% | | 42.5% | 33.3% | |
| 2-on-1 | 45.3% | 20.4% | | 41.5% | 17.8% | |
| 3-on-2 | 30.4% | 8.8% | 22.7 | 29.2% | 10.5% | 23.1 |
| 1-on-1 | 32.6% | 8.9% | | 27.0% | 6.2% | |
| 2-on-2 | 26.9% | 6.3% | | 24.3% | 6.2% | |
| 3-on-3 | 19.9% | 5.5% | 22.2 | 20.2% | 5.1% | 22.5 |
| 1-on-2 | 20.7% | 4.2% | | 18.0% | 4.6% | |
| 2-on-3 | 20.5% | 4.0% | | 19.5% | 3.3% | |
| dump-in | 1.1% | 8.5% | 21.3 | 0.8% | 9.7% | 21.6 |

Appendix Table 4 - Pace of play preceding controlled and dump-in entries for the AHL and SHL with percent of entries with a shot on goal and shooting percent of shots taken within 5 seconds of entry.

| | AHL | | | SHL | | |
|---|---|---|---|---|---|---|
| Speed Quintile | True Shooting % | Shot Distance (ft.) | Speed (ft/s) | True Shooting % | Shot Distance (ft.) | Speed (ft/s) |
| 1 | 3.15% | 39.8 | 9.5 | 2.68% | 38.6 | 10.3 |
| 2 | 3.27% | 41.4 | 18.5 | 2.93% | 40.8 | 18.7 |
| 3 | 3.43% | 42.3 | 23.3 | 3.22% | 41.5 | 23.8 |
| 4 | 3.99% | 40.8 | 27.8 | 3.30% | 40.3 | 28.8 |
| 5 | 4.32% | 39.8 | 42.1 | 3.42% | 39.1 | 42.8 |

Appendix Table 5 - Pace preceding a shot in the 2017-18 regular season for different leagues.

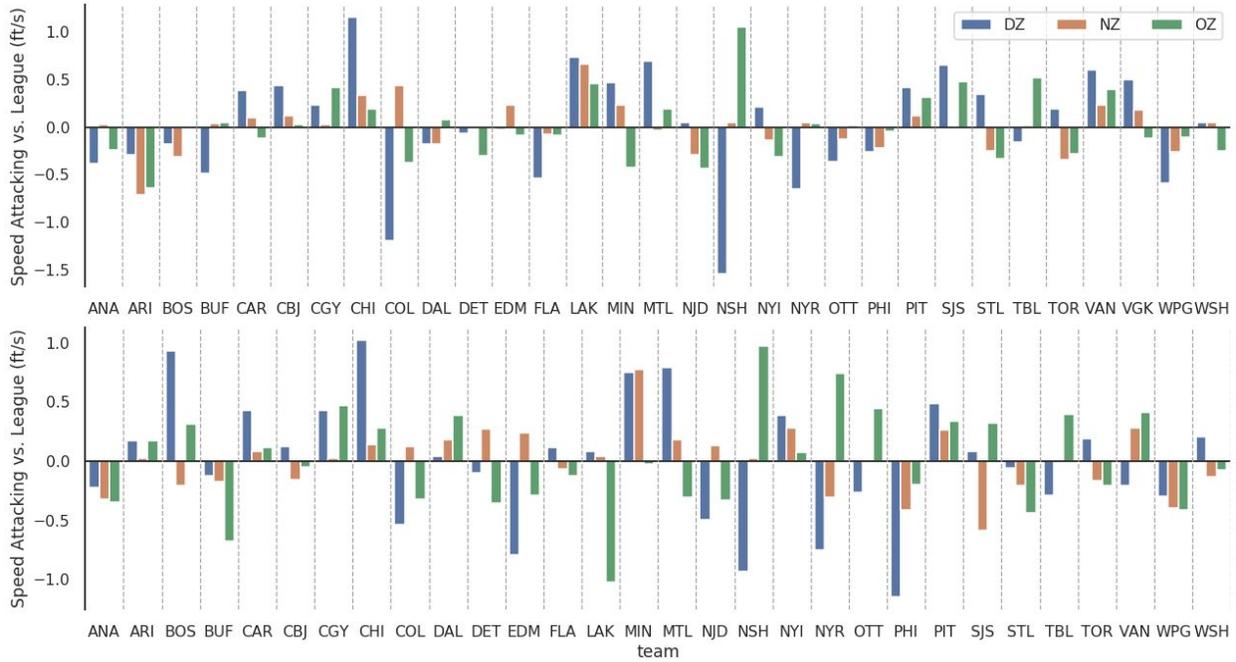

Appendix Figure 1 - Season over Season Differences in Total Speed for (Team Attacking vs. NHL Average). Top: 2017-18 Season Bottom: 2016-17 Season

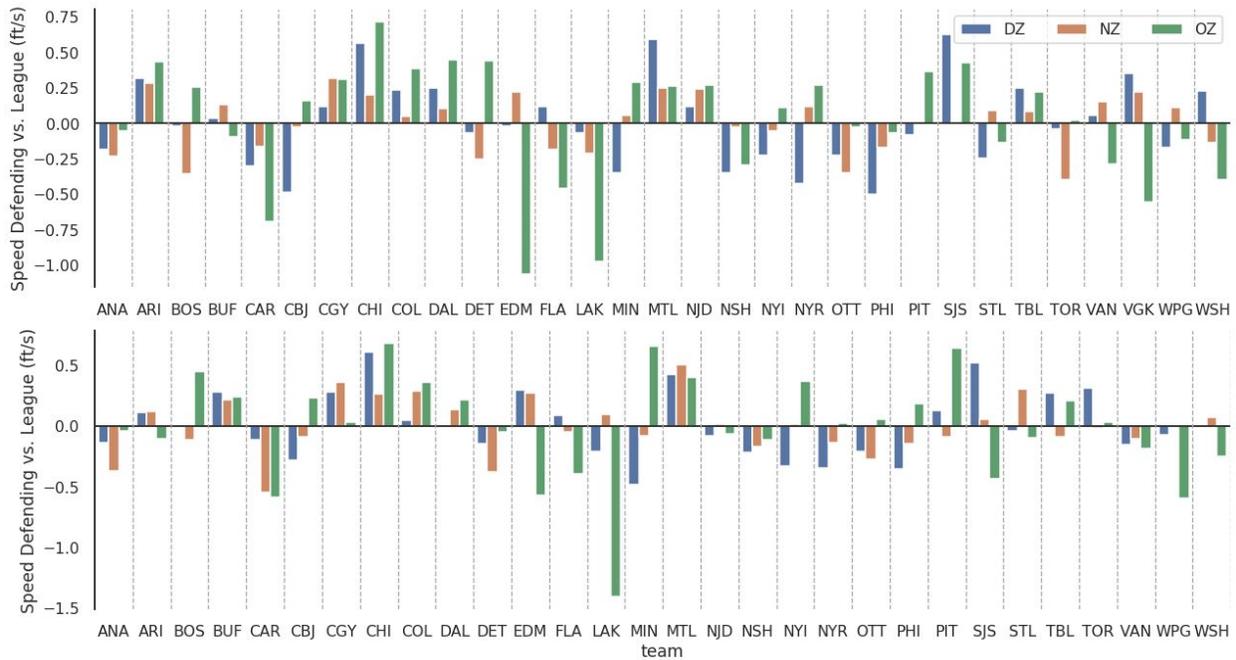

Appendix Figure 2 - Season over Season Differences in Total Speed for (Team Defending vs. NHL Average). Top: 2017-18 Season Bottom: 2016-17 Season

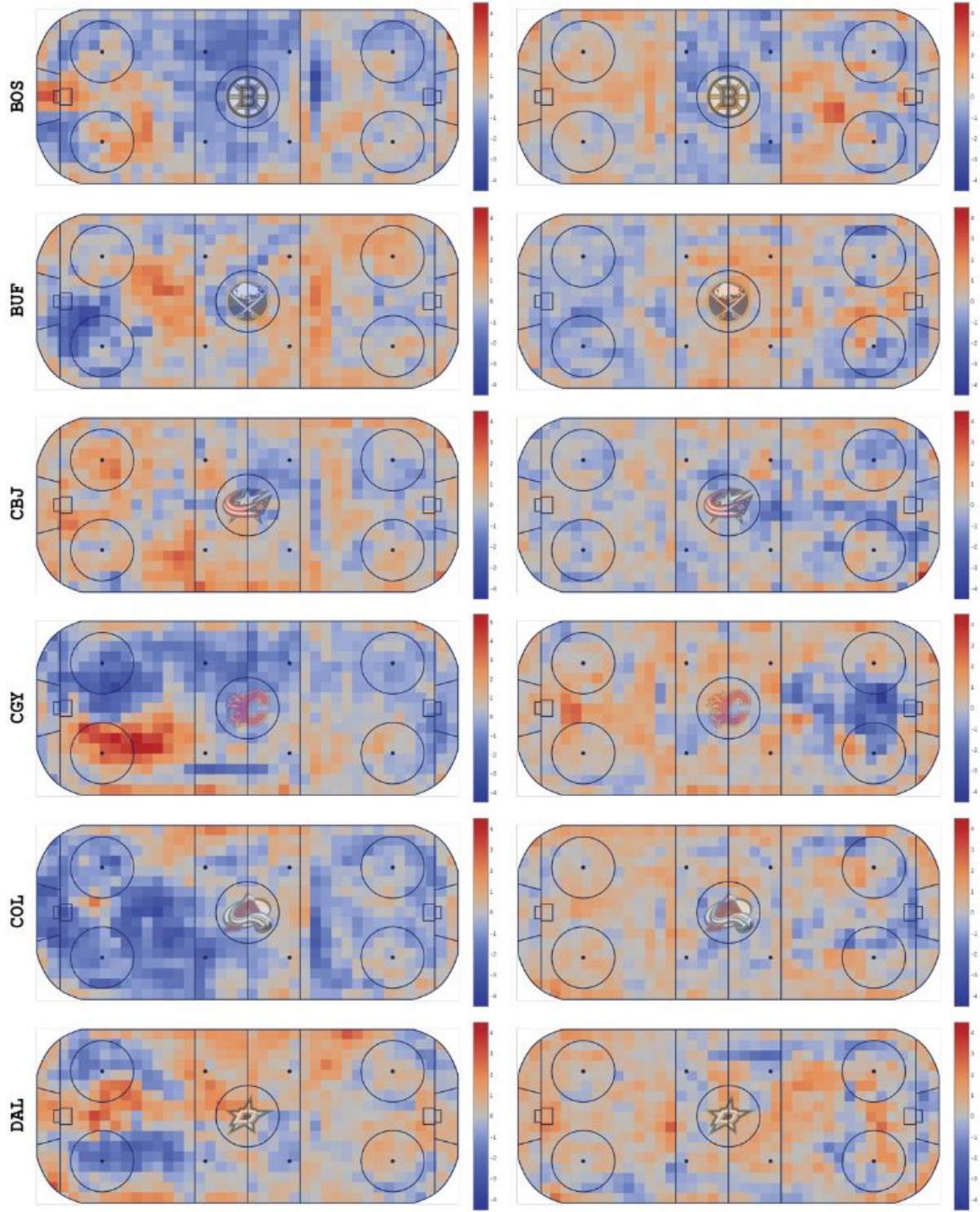

Appendix Figure 3 - Team-level $\phi_T$ polygrids for the remaining 23 NHL teams for the 2017-18 NHL Season. Attacking pace on right and defending pace on left. All units in ft/s.

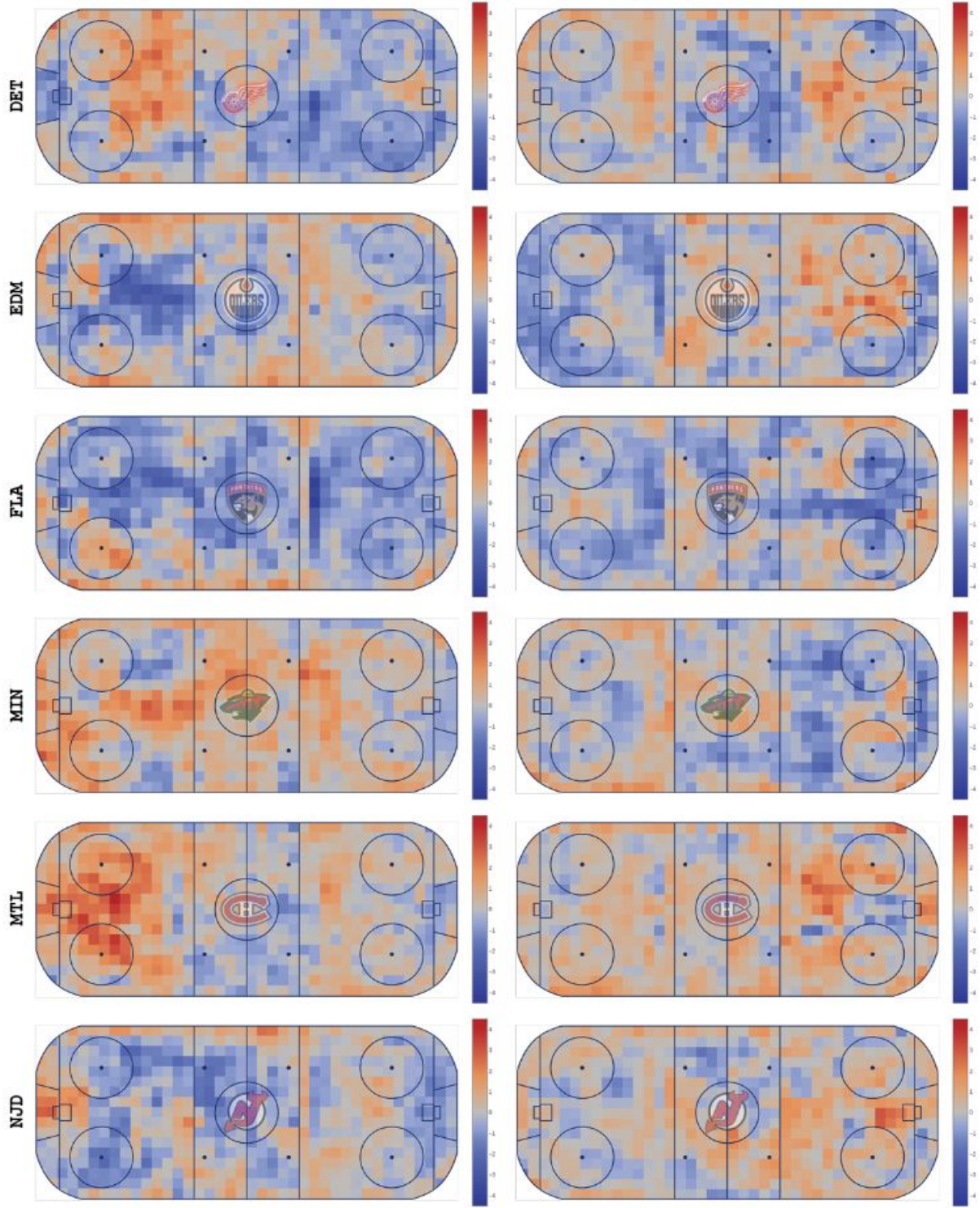

Appendix Figure 3 - Team-level $\phi_T$ polygrids for the remaining 23 NHL teams for the 2017-18 NHL Season. Attacking pace on right and defending pace on left. All units in ft/s.

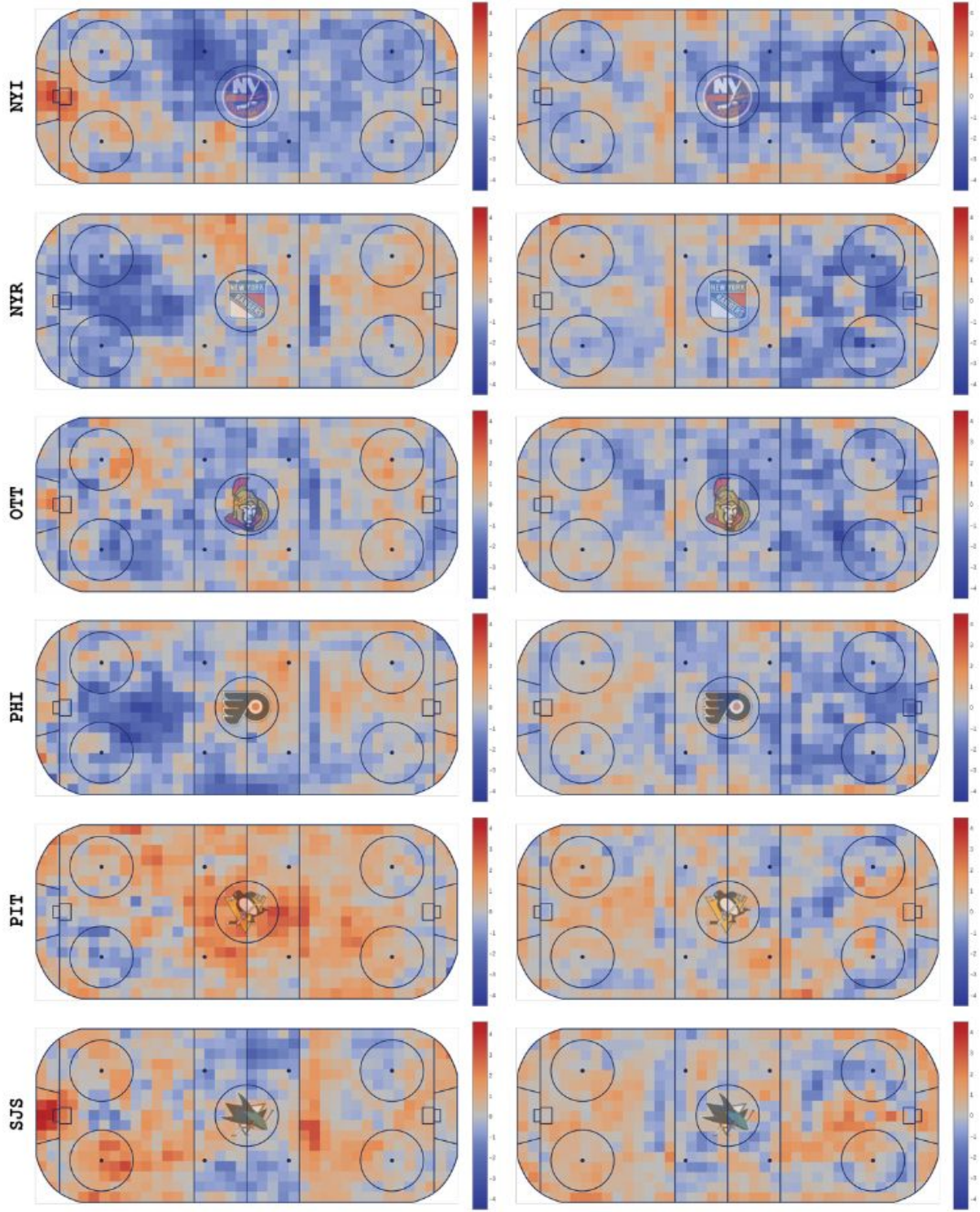

Appendix Figure 3 - Team-level $\phi_T$ polygrids for the remaining 23 NHL teams for the 2017-18 NHL Season. Attacking pace on right and defending pace on left. All units in ft/s.

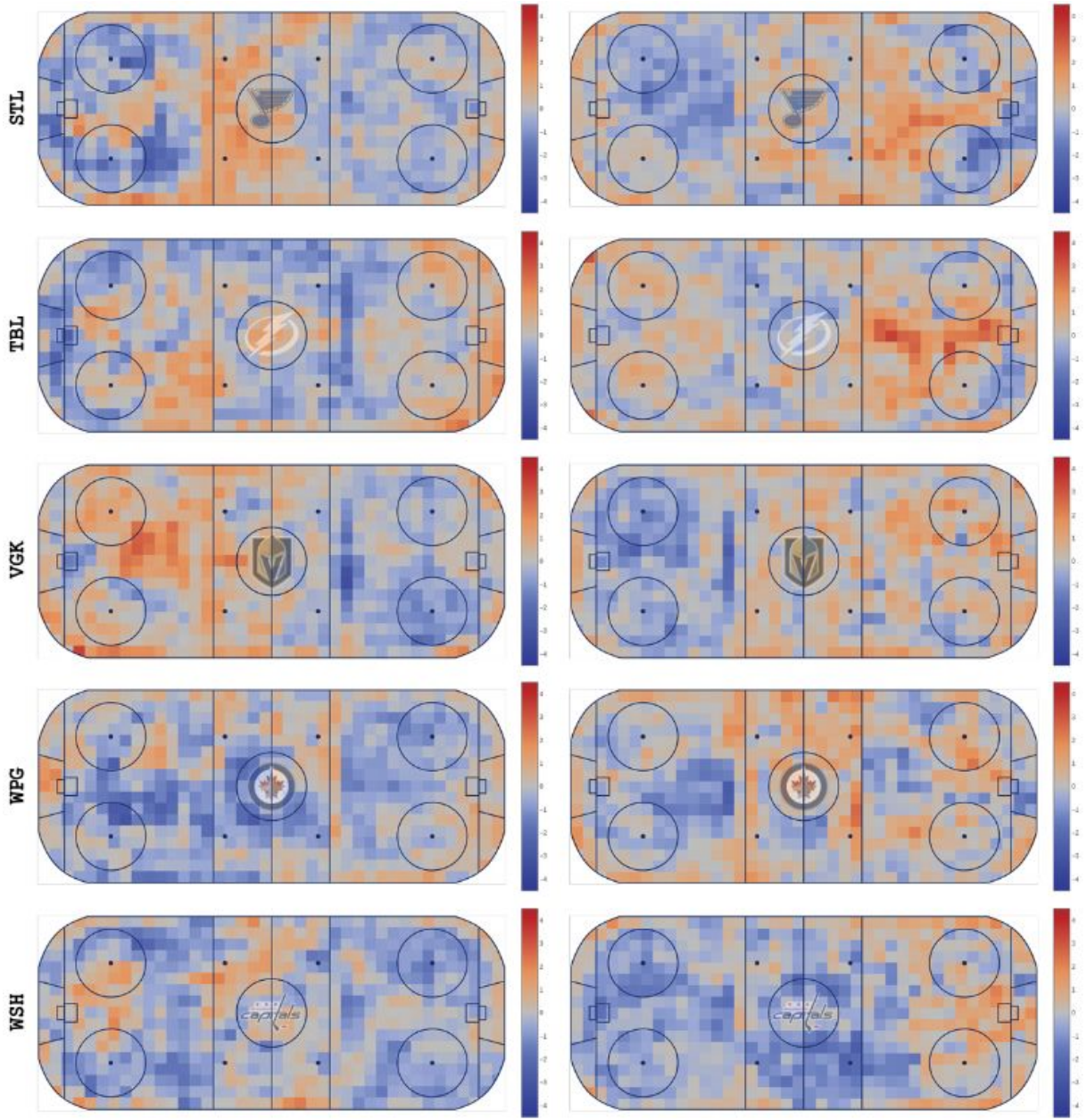

Appendix Figure 3 - Team-level $\phi_T$ polygrids for the remaining 23 NHL teams for the 2017-18 NHL Season. Attacking pace on right and defending pace on left. All units in ft/s.

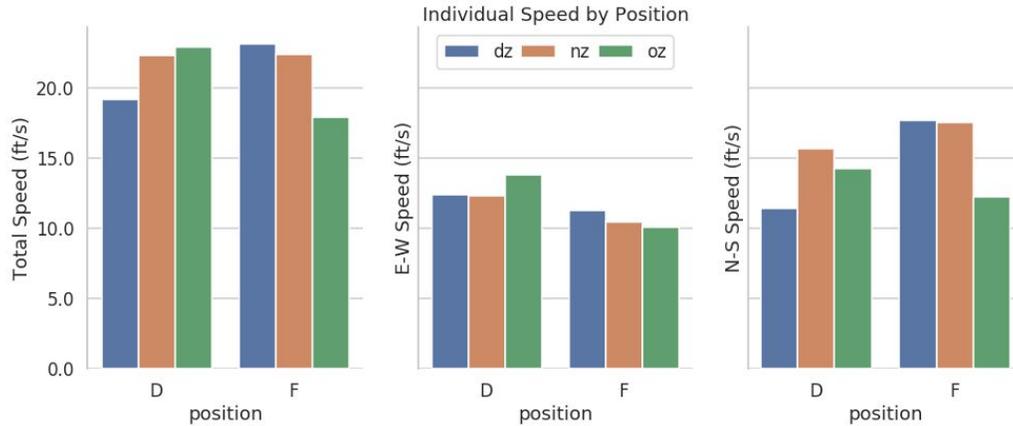

Appendix Figure 4 - Individual player pace by player position and zone.

Appendix Table 6 - Top 20 players by total speed WOWY % difference in the offensive zone

| team | player name | position | toi min | $\phi_T$ % | $\phi_{EW}$ % | $\phi_{NS}$ % | $\phi_N$ % |
|---|---|---|---|---|---|---|---|
| SJS | Joe Thornton | F | 643 | 9.0% | 15.9% | 3.4% | 3.4% |
| WPG | Paul Stastny | F | 256 | 8.0% | 11.1% | 7.0% | 9.3% |
| OTT | Mark Stone | F | 877 | 7.9% | 8.5% | 7.6% | 5.2% |
| VAN | Brendan Leipsic | F | 201 | 7.8% | 10.1% | 6.4% | 6.6% |
| PHI | Claude Giroux | F | 1,217 | 7.7% | 8.9% | 6.5% | 3.6% |
| OTT | Derick Brassard | F | 832 | 7.5% | 9.7% | 6.9% | 1.1% |
| PHI | Sean Couturier | F | 1,238 | 7.3% | 7.5% | 7.1% | 3.5% |
| BUF | Zach Bogosian | D | 303 | 6.7% | 9.6% | 4.4% | 9.8% |
| SJS | Joe Pavelski | F | 1,200 | 6.3% | 10.6% | 2.9% | 5.7% |
| EDM | Connor McDavid | F | 1,327 | 6.2% | 3.0% | 9.0% | 12.7% |
| ARI | Derek Stepan | F | 1,171 | 6.1% | 5.9% | 5.7% | 5.2% |
| DAL | Jamie Benn | F | 1,168 | 5.5% | 6.5% | 5.2% | 5.5% |
| BUF | Jason Pominville | F | 1,022 | 5.5% | 8.5% | 3.1% | 4.2% |
| VGK | Reilly Smith | F | 885 | 5.4% | 5.0% | 6.1% | 11.5% |
| WPG | Nikolaj Ehlers | F | 1,090 | 5.3% | 5.4% | 5.8% | 4.7% |
| DAL | Tyler Seguin | F | 1,219 | 5.3% | 6.8% | 4.4% | 3.6% |
| CGY | Micheal Ferland | F | 994 | 5.1% | 6.3% | 4.9% | 8.2% |
| SJS | Paul Martin | D | 204 | 5.1% | 3.7% | 6.1% | 2.5% |
| PHI | Travis Konecny | F | 1,036 | 5.0% | 2.9% | 6.3% | 8.0% |
| VGK | William Karlsson | F | 1,135 | 5.0% | 4.8% | 5.6% | 10.9% |

Appendix Table 7 - Bottom 20 players by total speed WOWY % difference in the offensive zone

| team | player name | position | toi min | $\phi_T$ % | $\phi_{EW}$ % | $\phi_{NS}$ % | $\phi_N$ % |
|---|---|---|---|---|---|---|---|
| PIT | Riley Sheahan | F | 868 | -7.0% | -7.8% | -6.3% | -6.1% |
| MTL | Byron Froese | F | 492 | -7.1% | -2.9% | -10.9% | -10.8% |
| COL | Nail Yakupov | F | 507 | -7.1% | -10.0% | -5.3% | -2.8% |
| MTL | Daniel Carr | F | 418 | -7.3% | -2.4% | -10.9% | -12.3% |

| team | player name | position | toi min | Φ_T % | Φ_EW % | Φ_NS % | Φ_N % |
|---|---|---|---|---|---|---|---|
| BUF | Jordan Nolan | F | 658 | -7.3% | -5.4% | -7.9% | -5.2% |
| PHI | Jordan Weal | F | 772 | -7.3% | -7.5% | -7.0% | -5.2% |
| TOR | Auston Matthews | F | 930 | -7.3% | -6.2% | -7.1% | -4.8% |
| DAL | Gemel Smith | F | 422 | -7.5% | -9.9% | -6.0% | -5.2% |
| BOS | Tim Schaller | F | 908 | -7.7% | -6.4% | -8.6% | -7.8% |
| ARI | Brad Richardson | F | 927 | -8.2% | -7.5% | -8.1% | -6.4% |
| VGK | Ryan Reaves | F | 206 | -8.2% | -7.7% | -8.1% | -12.1% |
| BOS | Noel Acciari | F | 676 | -8.3% | -7.9% | -7.8% | -7.3% |
| STL | Oskar Sundqvist | F | 377 | -8.4% | -7.0% | -9.6% | -5.3% |
| BUF | Jacob Josefson | F | 376 | -8.7% | -3.0% | -12.9% | -10.5% |
| VGK | P.E. Bellemare | F | 706 | -8.8% | -8.1% | -9.5% | -13.8% |
| PHI | Taylor Leier | F | 351 | -9.2% | -9.8% | -8.5% | -8.5% |
| CBJ | Mark Letestu | F | 206 | -9.6% | -8.7% | -9.7% | -1.2% |
| VGK | Tomas Nosek | F | 628 | -10.0% | -9.8% | -10.2% | -13.9% |
| VGK | William Carrier | F | 323 | -12.0% | -11.7% | -12.0% | -17.2% |
| CGY | Jaromir Jagr | F | 249 | -13.7% | -15.8% | -11.3% | -10.6% |

Appendix Table 8 - Top 20 players by total speed WOWY % difference in the neutral zone

| team | player name | position | toi min | Φ_T % | Φ_EW % | Φ_NS % | Φ_N % |
|---|---|---|---|---|---|---|---|
| TOR | Kasperi Kapanen | F | 377 | 5.2% | 6.5% | 4.6% | 3.6% |
| EDM | Brandon Davidson | D | 346 | 4.8% | 8.7% | 2.4% | 1.4% |
| WPG | Paul Stastny | F | 256 | 4.8% | 4.4% | 4.9% | 3.4% |
| TBL | J.T. Miller | F | 266 | 4.7% | 13.7% | -0.1% | 3.6% |
| VGK | Tomas Tatar | F | 250 | 4.6% | 6.4% | 4.5% | 4.6% |
| EDM | Adam Larsson | D | 1,169 | 4.6% | 6.9% | 3.3% | 5.5% |
| NSH | Kevin Fiala | F | 992 | 4.4% | 11.2% | 0.6% | 1.5% |
| EDM | Leon Draisaitl | F | 1,114 | 4.3% | 4.7% | 3.8% | 3.0% |
| BOS | Anders Bjork | F | 335 | 4.3% | 7.5% | 3.3% | 0.2% |
| PHI | Sean Couturier | F | 1,238 | 4.2% | 6.9% | 2.8% | 5.0% |
| NJD | Ben Lovejoy | D | 744 | 4.1% | 6.2% | 2.6% | 3.9% |
| NSH | Kyle Turris | F | 847 | 4.1% | 8.2% | 1.8% | 5.4% |
| CGY | Curtis Lazar | F | 605 | 3.9% | 2.6% | 4.5% | 8.7% |
| LAK | Kevin Gravel | D | 206 | 3.8% | 8.7% | 0.8% | -2.3% |
| PHI | Shayne Gostisbehere | D | 1,296 | 3.8% | 6.6% | 2.9% | 0.7% |
| OTT | Mike Hoffman | F | 1,165 | 3.8% | 10.9% | 0.6% | -2.6% |
| PHI | Claude Giroux | F | 1,217 | 3.6% | 7.4% | 2.0% | 3.2% |
| PHI | Ivan Provorov | D | 1,504 | 3.6% | 3.7% | 3.8% | 5.2% |
| CGY | Kris Versteeg | F | 234 | 3.5% | 6.5% | 2.5% | 2.1% |
| DET | Luke Witkowski | F | 209 | 3.5% | -0.4% | 5.1% | 9.8% |

Appendix Table 9 - Bottom 20 players by total speed WOWY % difference in the neutral zone

| team | player name | position | toi min | $\phi_T$ % | $\phi_{EW}$ % | $\phi_{NS}$ % | $\phi_N$ % |
|---|---|---|---|---|---|---|---|
| NSH | Austin Watson | F | 733 | -4.5% | -7.0% | -3.2% | -1.6% |
| VGK | P.E. Bellemare | F | 706 | -4.5% | -9.2% | -1.9% | -0.4% |
| NSH | Colton Sissons | F | 901 | -4.5% | -7.2% | -2.7% | -1.8% |
| CAR | Josh Jooris | F | 292 | -4.6% | -9.5% | -0.9% | -1.0% |
| TBL | Anthony Cirelli | F | 209 | -4.6% | -15.9% | 1.3% | 2.9% |
| ANA | Logan Shaw | F | 385 | -4.8% | -6.3% | -3.7% | -7.3% |
| DAL | Jason Dickinson | F | 227 | -4.8% | -10.6% | -1.9% | 1.5% |
| OTT | Alexandre Burrows | F | 664 | -4.8% | -9.3% | -3.1% | -4.4% |
| ARI | Jakob Chychrun | D | 862 | -5.0% | -3.8% | -5.3% | -4.8% |
| MIN | Zack Mitchell | F | 224 | -5.0% | -9.7% | -3.3% | -5.2% |
| STL | Vladimir Sobotka | F | 1,112 | -5.1% | -6.4% | -4.6% | -2.5% |
| VGK | Ryan Reaves | F | 206 | -5.1% | -10.7% | -2.9% | -2.2% |
| NYR | Neal Pionk | D | 508 | -5.1% | -9.1% | -3.1% | -5.1% |
| CGY | Sean Monahan | F | 1,012 | -5.2% | -4.7% | -5.3% | -5.9% |
| BUF | Josh Gorges | D | 439 | -5.2% | -8.4% | -3.3% | 0.5% |
| ARI | Brad Richardson | F | 927 | -5.3% | -7.4% | -4.1% | -1.8% |
| LAK | Tobias Rieder | F | 250 | -5.3% | -5.3% | -5.1% | -5.2% |
| LAK | Marian Gaborik | F | 357 | -5.7% | -9.0% | -4.3% | -5.3% |
| TBL | Adam Erne | F | 228 | -5.9% | -11.2% | -3.4% | -6.8% |
| NSH | Miikka Salomaki | F | 558 | -6.1% | -9.4% | -3.9% | -1.4% |

Appendix Table 10 - Top 20 players by total speed WOWY % difference in the defensive zone

| team | player name | position | toi min | $\phi_T$ % | $\phi_{EW}$ % | $\phi_{NS}$ % | $\phi_N$ % |
|---|---|---|---|---|---|---|---|
| NYR | Ryan Sproul | D | 243 | 7.2% | 8.8% | 6.4% | 6.0% |
| CHI | Erik Gustafsson | D | 574 | 6.8% | 7.6% | 6.9% | 8.4% |
| WPG | Joe Morrow | D | 248 | 6.7% | 6.6% | 7.7% | 5.3% |
| MIN | Jared Spurgeon | D | 1,100 | 6.1% | 6.4% | 6.0% | 6.4% |
| WPG | Toby Enstrom | D | 685 | 5.8% | 9.0% | 2.1% | -0.7% |
| MIN | Ryan Suter | D | 1,522 | 4.5% | 5.3% | 4.7% | 5.1% |
| WSH | Christian Djoos | D | 839 | 4.5% | 9.2% | 1.2% | 0.8% |
| LAK | Jeff Carter | F | 332 | 4.4% | 4.4% | 4.9% | 6.0% |
| TBL | Braydon Coburn | D | 966 | 4.4% | 5.5% | 2.9% | 3.5% |
| TBL | Andrej Sustr | D | 540 | 4.4% | 5.3% | 2.7% | 3.2% |
| PHI | Claude Giroux | F | 1,217 | 4.3% | 7.2% | 3.6% | 3.5% |
| WPG | Dustin Byfuglien | D | 1,296 | 4.1% | 4.0% | 4.0% | 2.6% |
| BUF | Marco Scandella | D | 1,487 | 4.1% | 5.3% | 2.2% | 2.3% |
| PHI | Sean Couturier | F | 1,238 | 3.8% | 6.5% | 3.3% | 3.2% |
| CBJ | M. Hannikainen | F | 256 | 3.5% | 1.6% | 4.2% | 3.7% |

| team | player name | position | toi min | ΦT % | ΦEW % | ΦNS % | ΦN % |
|---|---|---|---|---|---|---|---|
| STL | Alex Pietrangelo | D | 1,495 | 3.4% | 4.4% | 2.6% | 3.4% |
| OTT | Ben Harpur | D | 535 | 3.4% | 2.5% | 4.5% | 5.7% |
| BOS | David Pastrnak | F | 1,149 | 3.4% | 3.0% | 3.9% | 3.9% |
| COL | Mark Barberio | D | 664 | 3.3% | 1.1% | 4.3% | 4.0% |
| TOR | Morgan Rielly | D | 1,297 | 3.1% | 1.5% | 4.5% | 5.2% |

Appendix Table 11 - Bottom 20 players by total speed WOWY % difference in the defensive zone

| team | player name | position | toi min | ΦT % | ΦEW % | ΦNS % | ΦN % |
|---|---|---|---|---|---|---|---|
| TOR | Dominic Moore | F | 448 | -7.4% | -4.8% | -9.7% | -11.5% |
| NSH | Yannick Weber | D | 534 | -7.5% | -5.7% | -9.5% | -11.1% |
| CGY | Jaromir Jagr | F | 249 | -7.7% | -5.4% | -7.9% | -4.8% |
| PIT | Greg McKegg | F | 204 | -7.7% | -10.3% | -6.2% | -4.9% |
| BUF | Nicholas Baptiste | F | 292 | -7.8% | -9.3% | -7.2% | -7.9% |
| PHI | Jordan Weal | F | 772 | -7.9% | -7.1% | -9.3% | -10.5% |
| BUF | Kyle Okposo | F | 974 | -7.9% | -8.6% | -7.4% | -7.6% |
| DAL | Jason Dickinson | F | 227 | -7.9% | -5.8% | -10.3% | -10.2% |
| TBL | Anton Stralman | D | 1,370 | -8.0% | -9.5% | -6.5% | -8.0% |
| PHI | Dale Weise | F | 465 | -8.2% | -6.6% | -10.3% | -13.6% |
| FLA | Maxim Mamin | F | 268 | -8.7% | -5.0% | -11.7% | -13.3% |
| CGY | Kris Versteeg | F | 234 | -9.0% | -8.2% | -8.3% | -8.1% |
| COL | Samuel Girard | D | 1,013 | -9.0% | -10.9% | -7.6% | -7.7% |
| VAN | A.Burmistrov | F | 248 | -9.1% | -9.0% | -9.3% | -10.6% |
| NSH | Miikka Salomaki | F | 558 | -9.2% | -7.4% | -10.7% | -13.1% |
| NSH | Ryan Hartman | F | 257 | -9.5% | -8.3% | -10.5% | -13.3% |
| EDM | Eric Gryba | D | 278 | -9.6% | -12.9% | -6.1% | -4.1% |
| NSH | Austin Watson | F | 733 | -9.7% | -8.6% | -10.8% | -13.3% |
| NSH | Colton Sissons | F | 901 | -9.9% | -8.8% | -10.9% | -13.0% |
| TBL | Anthony Cirelli | F | 209 | -10.2% | -16.0% | -6.7% | -8.8% |